\documentclass[fleqn,usenatbib]{mnras}

\usepackage{newtxtext,newtxmath}
\usepackage{amsmath}

\usepackage[T1]{fontenc}

\DeclareRobustCommand{\VAN}[3]{#2}
\let\VANthebibliography\thebibliography
\def\thebibliography{\DeclareRobustCommand{\VAN}[3]{##3}\VANthebibliography}

\usepackage{graphicx}	
\usepackage{amsmath}	

\newcommand\software[1]{\textsc{#1}}

\newcommand\scipy{\software{SciPy}}

\newcommand{\tw}[1]{\textcolor{black}{#1}}

\newcommand{\Del}[1]{}

\newcommand{\tamb}{$T_{\text{amb}}$}
\newcommand{\tspot}{$T_{\text{spot}}$}
\newcommand{\xspot}{$x_{\text{spot}}$}
\newcommand{\fspot}{$f_{\text{spot}}$}
\newcommand{\teff}{$T_{\text{eff}}$}
\newcommand{\vsini}{$v \sin{i}$}
\newcommand{\logg}{$\log{g}$}

\newcommand{\rev}[1]{#1}

\usepackage{xcolor} 
\definecolor{aureolin}{rgb}{0.99, 0.93, 0.0}

\usepackage[math]{cellspace}

\title[Star spots affect measured metallicity]{Stellar spots cause measurable variations in atmospheric metallicity}

\author[Tanner A. Wilson and Andrew R. Casey]{
Tanner A. Wilson,$^{1,2}$\thanks{Corresponding Author E-mail: tanner.wilson@monash.edu}
and
Andrew R. Casey,$^{1,2}$
\\
$^{1}$School of Physics \& Astronomy, Monash University, Victoria, Australia\\
$^{2}$Center of Excellence for Astrophysics in Three Dimensions (ASTRO-3D)\\
}

\date{Accepted XXX. Received YYY; in original form ZZZ}

\pubyear{2023}

\begin{document}
\label{firstpage}
\pagerange{\pageref{firstpage}--\pageref{lastpage}}
\maketitle

\begin{abstract}
\tw{To accurately measure a star's atmospheric parameters and chemical abundances, it is crucial to have high-quality spectra. Analysing the detailed chemical abundances of groups of stars can help us better understand nucleosynthesis, galactic chemical enrichment, and stellar evolution. In this study, we explored whether stellar spots can affect a star's inferred metallicity and, if so, where the impact is the strongest. To investigate this, we created synthetic infrared spectra that included stellar spots for a sample of main-sequence stars \rev{younger than the sun}. We then applied two models to the data: one that accounted for spots and one that did not. From this, we can determine the bias introduced when fitting spotted spectra with a non-spotted model and how this bias varies with different parameters. Our findings revealed that fitting spotted spectra with a non-spotted model can introduce a scatter of up to 0.05 dex in the inferred metallicity, especially for stars with high levels of spot coverage. This bias is similar in magnitude to other relevant effects, such as atomic diffusion, radiative levitation, or non-local thermodynamic equilibrium. We also found that the effect is most pronounced in young stars and decreases with age. These results suggest that stellar spots can introduce a systematic uncertainty in metallicity that is not currently accounted for in spectroscopic analysis. This could potentially limit scientific inferences for population-level studies \rev{of young stars} and differential abundance analyses.}
\end{abstract}

\maketitle

\section{Introduction}
\label{sec:intro}

It is widely assumed that the elemental abundances in a star's atmosphere accurately reflect the abundances of the material from which the star formed \citep{gibson_galactic_2003,pagel_nucleosynthesis_2009,salaris_chemical_2017}. This assumption is critical for chemical tagging \citep{anders_galactic_2016, randich_gaia-eso_2022}, understanding galactic formation \citep{gibson_galactic_2003}, and the synthesis of elements across cosmic time \citep{mcwilliam_origin_2004,johnson_origin_2020}. Precise 
measurements of elemental abundances are essential in many areas of astrophysics. For example, chemical tagging allows us to track the history of the galaxy, which would be impossible with biased measures of abundance. Differential abundance techniques \citep{onehag_m67-1194_2011, melendez_18_2014, reggiani_first_2016, maia_revisiting_2019, liu_detailed_2020, nissen_high-precision_2020, spina_chemical_2021} -- employed for solar twins and planet-hosting stars -- claim very precise abundance measurements, which are essential for probing planet formation \citep{tayar_guide_2022}. Similarly, when determining cluster ages \citep{bensby_possible_2004, pont_isochrone_2004}, the turn-off age of a star is particularly useful for this purpose because a small change in colour/magnitude, which depends on metallicity, indicates a relatively large change in age compared to the main-sequence.

Recognising that surface abundances may change over a star's evolution is important. The surface abundances can change due to numerous processes. Atomic diffusion and radiative levitation introduce surface abundance variations on the scale of 0.05 dex, with a magnitude and bias that depends on the element and the stellar effective temperature \citep{onehag_abundances_2014}. Enhanced mixing can also cycle material to the surface. Nuclear reactions, such as lithium depletion \citep{pinsonneault_stellar_2002} or CNO cycling \citep{crowther_physical_2007}) enhance and deplete specific surface abundances and isotopic ratios. Accretion can enhance surface metallicity and vary particular elemental abundances for a short time depending on the companion type \citep{pasquini_evolved_2007, maldonado_connecting_2019, laughlin_possible_1997}. For example, mass loss can strip away H-rich surface regions in Wolf-Rayet stars \citep{crowther_physical_2007} - increasing the observed stellar metallicity or carrying away surface metals which will have a small to negligible decrease of surface metals on the main sequence. 

These effects are usually ignored when estimating a star's stellar parameters and chemical abundances. Most spectroscopic analyses usually adopt some simplifying assumptions to make the computation time tractable.
For example, we usually assume the stellar photosphere can be represented in one dimension (1D) and that baryonic matter can be described by thermal distributions in small regions (local thermal equilibrium; LTE). These assumptions can particularly influence the measured stellar parameters \citep[e.g.,][]{blanco-cuaresma_modern_2019}. Both can lead to an over-estimate of the temperature gradient in the atmosphere and an underestimation of the density, which can result in an over-estimate of the abundance. We also typically ignore magnetic activity, but recently \citet{spina_how_2020} showed it has a measurable impact on the chemical abundances of young, fast-rotating stars.
While these assumptions simplify inference, it is important to consider their effects when reaching conclusions.

Stars have spots, which are important indicators of the rotational rate of stars, especially along the main sequence \citep{mcquillan_rotation_2014, santos_surface_2021}. The properties of stellar spots \rev{and their effect on the observed properties of a star} vary with: age, rotation rate, mass, and metallicity \citep{mathur_magnetic_2014, karoff_influence_2018, nichols-fleming_determination_2020}.
\rev{For example, as the rotation rates of stars decrease with age, the average magnetic activity likewise decreases. This results in smaller short lived spots that cover only a small fraction of the stellar surface \citep{cao_starspots_2022}.}
Spot properties can be generalised by: their coverage across the stellar surface, the temperature difference relative to the surrounding, and the occurrence pattern.
\citet{cao_starspots_2022} recently quantified the spot parameters of stars in the Pleiades and M67. They found that young or fast rotating stars tend to be more magnetically active and have a greater spot coverage than their older, slower counterparts. 

The spotted areas of the star can be thousands of degrees cooler than the surrounding areas \rev{- solar spots for example can be 500-2000K cooler than the surrounding photosphere \citep{berdyugina_starspots_2005,herbst_starspots_2020}}. A spotted star's stellar spectra are more complex than their non-spotted counterparts \citep{morris_stellar_2019}. Accurate inference of stellar parameters requires a model that reflects the stellar spectra well. In this work, we quantify the effect of fitting spotted spectra with non-spotted models and identify the parts of the main sequence where the effect is most prevalent.
In Section \ref{sec:methods}, we outline the generative model for stellar spectra with spots and describe our choices of stellar parameters before outlining the fitting procedure used.
In Section \ref{sec:results}, we present the difference in the recovered stellar parameters with the spotted and non-spotted models. We discuss parts of the main sequence where the effect is most prevalent. Finally, in Section \ref{sec:discussion}, we place those results in the context of other significant effects on measured stellar metallicity and provide recommendations for high-precision spectroscopic investigations in specific regions of stellar evolution.

\section{Method}
\label{sec:methods}

\subsection{Stellar parameters for a population of fake stars}
\label{sec:stellar_parameters}
We prepare a sample of stellar spectra that spans the main sequence to estimate the impact that stellar spots can have on the accuracy of inferred stellar parameters. This sample is intended to be indicative of a possible population of main-sequence stars but not intended to represent which stars would, or would not, have spots.
We generate 1500 spectra of main-sequence and early post-main-sequence stars with various values of mass, age, metallicity, \vsini, \fspot, and \xspot\ across the HR diagram.
We drew masses from a Salpeter initial mass function \citep{salpeter_luminosity_1955} between 0.5 and 1.5 $M_\odot$ with $\alpha = 2.35$. This limits our range of masses to those with a radiative surface and convective core and reaches beyond the Kraft break \citep{kraft_studies_1967}. Metallicity is drawn from a distribution to approximately reflect what is observed in the Milky Way. Specifically, we defined a variable $\phi$ to be drawn from a Beta distribution
\begin{equation}
    \phi \sim \mathcal{B}\left(\alpha=10, \beta=2\right)
\end{equation}
and applied a transform from $\phi$ to [Fe/H] by requiring the metallicities be bounded between $[\mathrm{Fe/H}]_\mathrm{min} =-2$ and $[\mathrm{Fe/H}]_\mathrm{max} = +0.5$. We also required that the mode of $\phi$, defined as $\frac{\alpha - 1}{\alpha + \beta - 2}$ for a Beta distribution, occurs at Solar metallicity. This leads to the transform:
\begin{equation}
    [\mathrm{Fe/H}] = \left(\frac{}{}[\mathrm{Fe/H}]_\mathrm{max}-[\mathrm{Fe/H}]_\mathrm{min}\right)\left(\phi - \frac{\alpha - 1}{\alpha + \beta - 2}\right) \quad .
\end{equation}

The stars we generate mock data for in this work span from the zero-age main sequence (ZAMS) to low-luminosity subgiants. We draw equivalent evolutionary phase (EEP) values from a uniform distribution EEP $\sim \mathcal{U}(200,450)$, where $\mathcal{U}\left(x,y\right)$ denotes a uniform prior between x and y.
The bounds of this range (200 and 450) represent the ZAMS and the low-luminosity subgiant phase, respectively. 
Using the EEP, mass and metallicity, we interpolate a position along the MIST stellar isochrones \citep{morton_isochrones_2015} to calculate the expected \teff\ and \logg\ for each random star. We also obtain the star's age (post-ZAMS) that we can use in conjunction with the other stellar parameters to determine rotational properties (see below). We have limited the age of the stars we consider in this work up to the age of the Sun. This is the range available for rotational rate and convective turnover timescales from the sources we draw from in this work. This limits the post-MS stars we consider to more massive stars. We briefly discuss bias' which may introduce in Section \ref{sec:discussion}.

The surface rotation period is interpolated from stellar cluster-tuned rotational isochrones given the stellar age and mass (Table A1 in \citet{spada_angular_2016}). Rotational broadening \vsini\ can then be calculated by combining the rotational period, the radius from the interpolated isochrone model, and an inclination angle. We have drawn inclination from a uniform distribution in $\cos{i} \sim \mathcal{U}(0, 1)$.

\fspot\ is related to the Rossby number, $R_o$, which is defined as the ratio of the surface rotational period to the convective turnover timescale ($\tau_{\textrm{conv}}$). $\tau_{\textrm{conv}}$ is interpolated from Table 1 in \citet{landin_theoretical_2010} given the stellar age and mass. Combining this value with the rotational period, we obtain $R_o$. \fspot\ is then calculated from the relationship between \fspot\ and $Ro$ identified in \citet{cao_starspots_2022} (Eq. 5):
\begin{equation}
f_{\text{spot}} = 
\left\{
    \begin{array}{lr}
        0.278, & \log{R_o} \leq -0.729 \\
        0.0623\ R_o^{-0.881}, & \log{R_o} > -0.729
    \end{array}
\right\}.
\end{equation}

There is some scatter in \fspot\, which is not accounted for by this relation (see left panel of Figure 7 in \citet{cao_starspots_2022}). For this reason, we add random noise to our calculated \fspot\, which is drawn from a normal distribution with a standard deviation of 0.1
it is clear that there is some scatter in \fspot\ About 

We assume \xspot\ is drawn from a uniform distribution $x_{\text{spot}} \sim \mathcal{U}(0.8, 1.0)$. This represents the limits set when fitting \xspot\ in \citep{cao_starspots_2022}, which is motivated by temperature bounds which they discuss in more detail in Section 2.2. \xspot\ does not appear to have a clear relationship with other stellar parameters, but it - and \fspot\ - may vary on multiple periodic timescales as they do for the Sun. The stochastic nature of stellar observations - and the admittedly simple nature of the model - means that \fspot\ and \xspot\ are random draws from the possible stellar spot parameters. We will eventually find that \xspot\ has little effect on the bias introduced by fitting spotted spectra with a non-spotted model, so move forward with the knowledge that we have good coverage when modelling over the range of possible parameters.

\begin{figure}
    \includegraphics[width=0.5\textwidth]{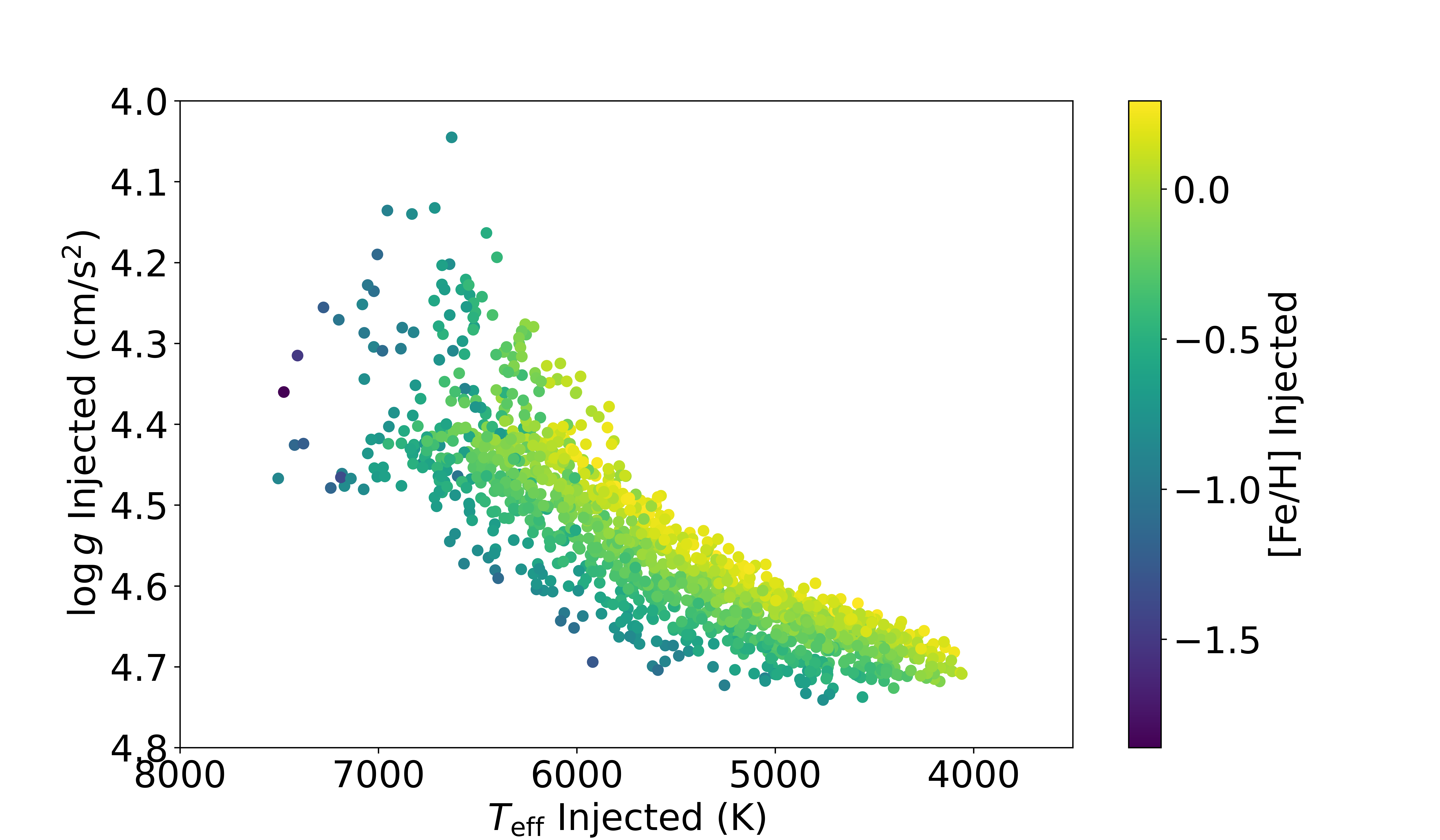}
    \caption{HR diagram of the 1500 sets of stellar parameters drawn from physically motivated distributions of mass, metallicity and age coloured by [Fe/H].}
    \label{fig:HR}
\end{figure}

\subsection{Spotted spectrum generative model}

\label{sec:gen}
We build upon the work of \citep{cao_starspots_2022}, where a forward model is developed to model the effect of starspots and to estimate the fractional spot coverage of stars in the Pleiades and M67. Their model assumes that the spectrum of a spotted star can be broken into spotted and non-spotted components. These two components have the same \logg\, [Fe/H], microturbulent velocity, and the same surface rotational velocity (\vsini), but the two components vary in temperature. The spot and ambient temperatures (\tspot\ and \tamb) are related by \tspot\  = \xspot \tamb, and are coupled to the effective temperature of the star following the approach of \citet{somers_impact_2015}
\begin{equation}
    T_{\text{eff}} = T_{\text{amb}}(1 - f_{\text{spot}} + f_{\text{spot}}x_{\text{spot}}^{4} )^{\frac{1}{4}},
    \label{eq:teff}
\end{equation}
where \fspot is the fractional surface spot coverage. From these relations, the set \{\teff, \xspot,\fspot\} define a pair of ambient and spot temperatures that preserve stellar luminosity.

We calculated a grid of synthetic spectra, which we interpolate between to generate the predicted spectra for a spotted or non-spotted model. The list of atomic and molecular transitions is from \citep{shetrone_sdss-iii_2015, smith_apogee_2021}. We used a grid of plane-parallel MARCS \citep{gustafsson_grid_2008} model photospheres that span dimensions in effective temperature, surface gravity, and metallicity.\footnote{We calculated spectra using spherical models as well, but in practice, only spectra from plane-parallel models (i.e., main-sequence stars) are used in this work.} Microturbulence was \rev{kept fixed at 1.15\,km\,s$^{-1}$} for main-sequence stars and we assumed that [$\alpha$/H] scales with [Fe/H] (i.e., the so-called `standard' composition in MARCS). The abundance dimensions [C/M] and [N/M] were kept fixed at zero. We used Korg \citep{wheeler_korg_2022} to synthesise all model spectra at high resolution, which we then convolved and down-sampled to match the (uniform in log) pixel spacing used in the APOGEE data reduction pipeline \citep{holtzman_apogee_2018}. The convolution kernel includes two components that enter multiplicatively: one assuming a constant spectral resolution $R = \lambda/\Delta\lambda$ of 22,500, and another representing rotational broadening $v\sin{i}$. We convolved each spectrum with a grid of $v\sin{i}$ values that were uniformly spaced in $\log{v\sin{i}}$ from 0-100 km\,s$^{-1}$ in order to match the setup for the APOGEE analysis pipeline. Naturally, for low $v\sin{i}$ values, the line spread function of the instrument will dominate.

With this grid of spectra and some given spectral parameters $\{T_\mathrm{eff},\log{g},[\mathrm{Fe/H}],\log{v\sin{i}}\}$, we interpolate the spotted and ambient spectra and combine them in a fractional manner with wavelength as if they were separate black-body spectra in order to produce a flux-preserving combined spectrum:
\begin{equation}
    B(T_{\text{eff}} , \lambda) \ = \ f_{\text{spot}}B(T_{\text{spot}}, \lambda) + (1 - f_{\text{spot}}) B(T_{\text{amb}}, \lambda) \quad .
\end{equation}

In total, our forward model for predicting spotted spectra includes six parameters: $T_\mathrm{eff}$, $\log{g}$, [Fe/H], $\log{v\sin{i}}$, \xspot, and \fspot. This model is equally capable of predicting non-spotted spectra by fixing \fspot\ to zero or \xspot\ to unity.

\tw{Using the 1500 sets of parameters outlined in Section \ref{sec:stellar_parameters} we generated synthetic spotted stellar
spectra.
We also apply realistic noise at each pixel from a Gaussian distribution with standard deviation = 0.01, assuming a signal-to-noise ratio of 100. Continuum normalisation is performed by assuming a running mean of the spectra, and during fitting, this procedure is applied to the fake spectrum (data) and to the model spectrum.}

We now have the tools to determine the effect of fitting spotted spectra with non-spotted models. We do this by finding the best-fitting stellar parameters given the synthetic spectra fitted twice: first with the model described in Section \ref{sec:gen} and then with a non-spotted model (e.g., \fspot\ fixed at 0 and \xspot\ fixed at 1).
\tw{Here we have performed least-squares fitting through the Levenberg-Marquardt algorithm implemented in \scipy.} We found that fitting the spotted parameters can be non-trivial. The likelihood surfaces are multimodal and degenerate, requiring informed choices about the initialisation of fitting. 
To resolve this issue, we performed a coarse evaluation of parameters (on a grid) before starting optimisation.

\section{Results}
\label{sec:results}

\begin{figure*}
    \centering
    \includegraphics[width = \textwidth]{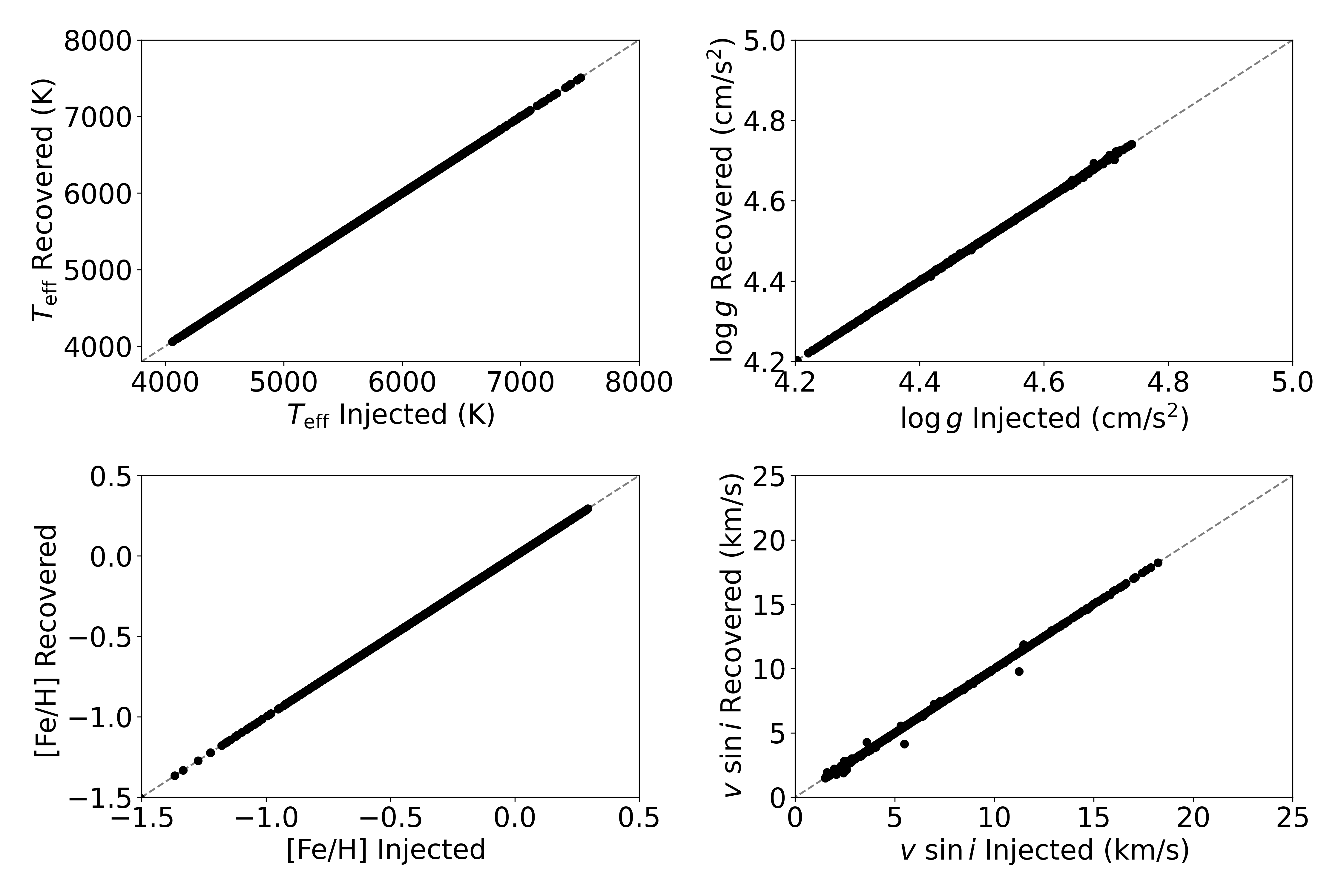}
    \caption{\tw{Recovered traditional stellar parameters (\teff, \logg, [Fe/H] and \vsini) from fitting synthetic spotted spectra with a spotted model of the stellar atmosphere against the corresponding injected parameters. We consistently accurately recover each injected value when a spotted model of the stellar atmosphere is employed to fit the spotted synthetic spectra.}}
    \label{fig:recov_test}
\end{figure*}

\begin{figure}
    \includegraphics[width=0.5\textwidth]{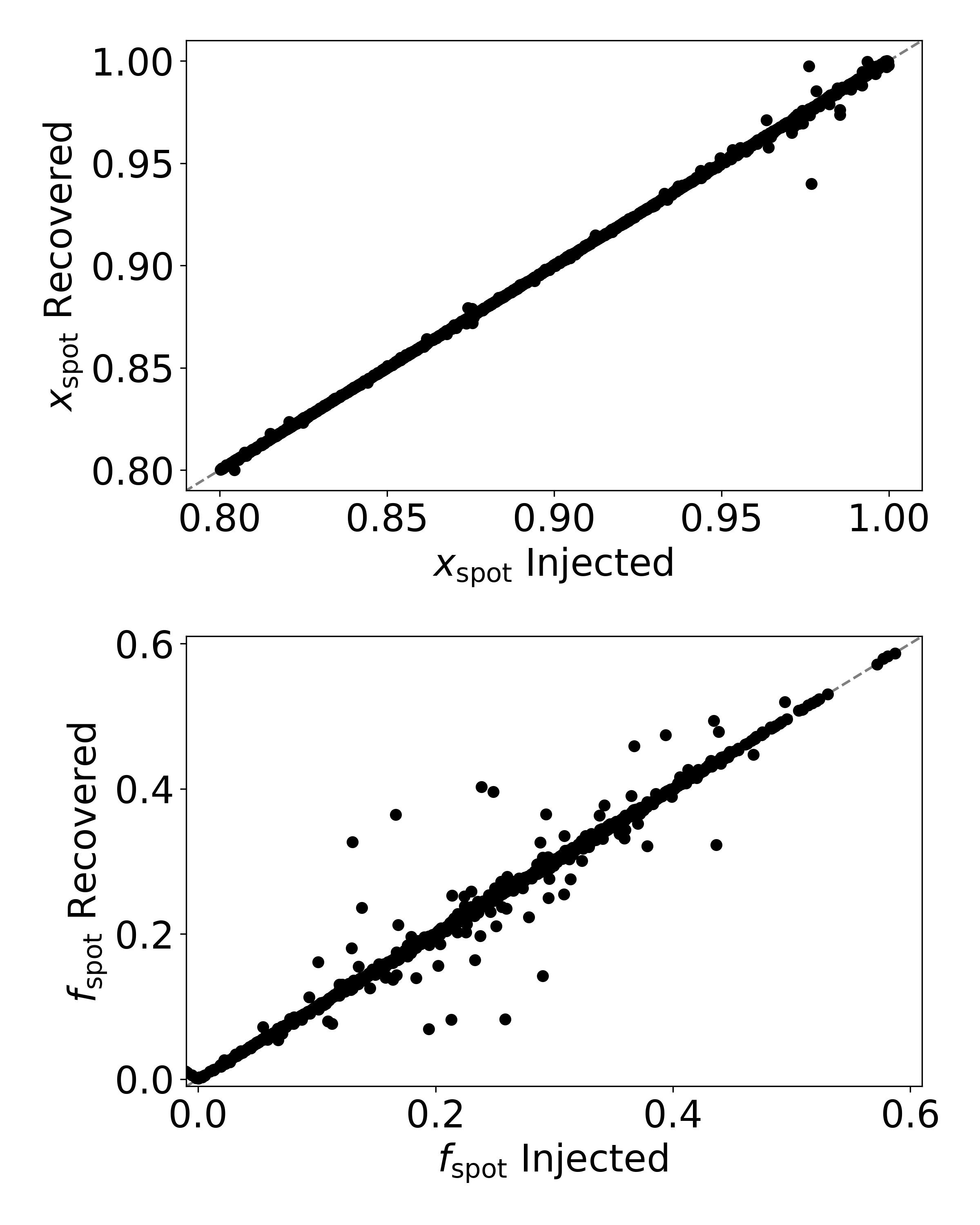}
    \caption{\tw{Recovered spot parameters (\xspot\ and \fspot) from the synthetic spotted spectra fitted with a spotted model of the stellar spectra against the injected parameters of the synthetic spectra. We identify that the spot parameters are not always accurately recovered through the fitting procedure. The recovered spot parameters are notably more inaccurate as \xspot \ approaches 1. }}
    \label{fig:recov_spot_params}
\end{figure}

\tw{We began by confirming that we could accurately recover the injected parameters. 
The best-fit parameters following fitting the synthetic spotted spectra with a spotted model are shown in Figures \ref{fig:recov_test} and \ref{fig:recov_spot_params}. \teff, \logg, [Fe/H], and \vsini\ are recovered accurately for every injected parameter set. While we identify scatter in recovered \fspot\, this appears not to affect the accuracy of the recovery of the traditional stellar parameters. We move forward confident that any difference in the recovered parameters between fitting with the spotted and non-spotted models results from the model differences rather than the fitting procedure employed in this work.}

\begin{figure*}
    \includegraphics[width=\textwidth]{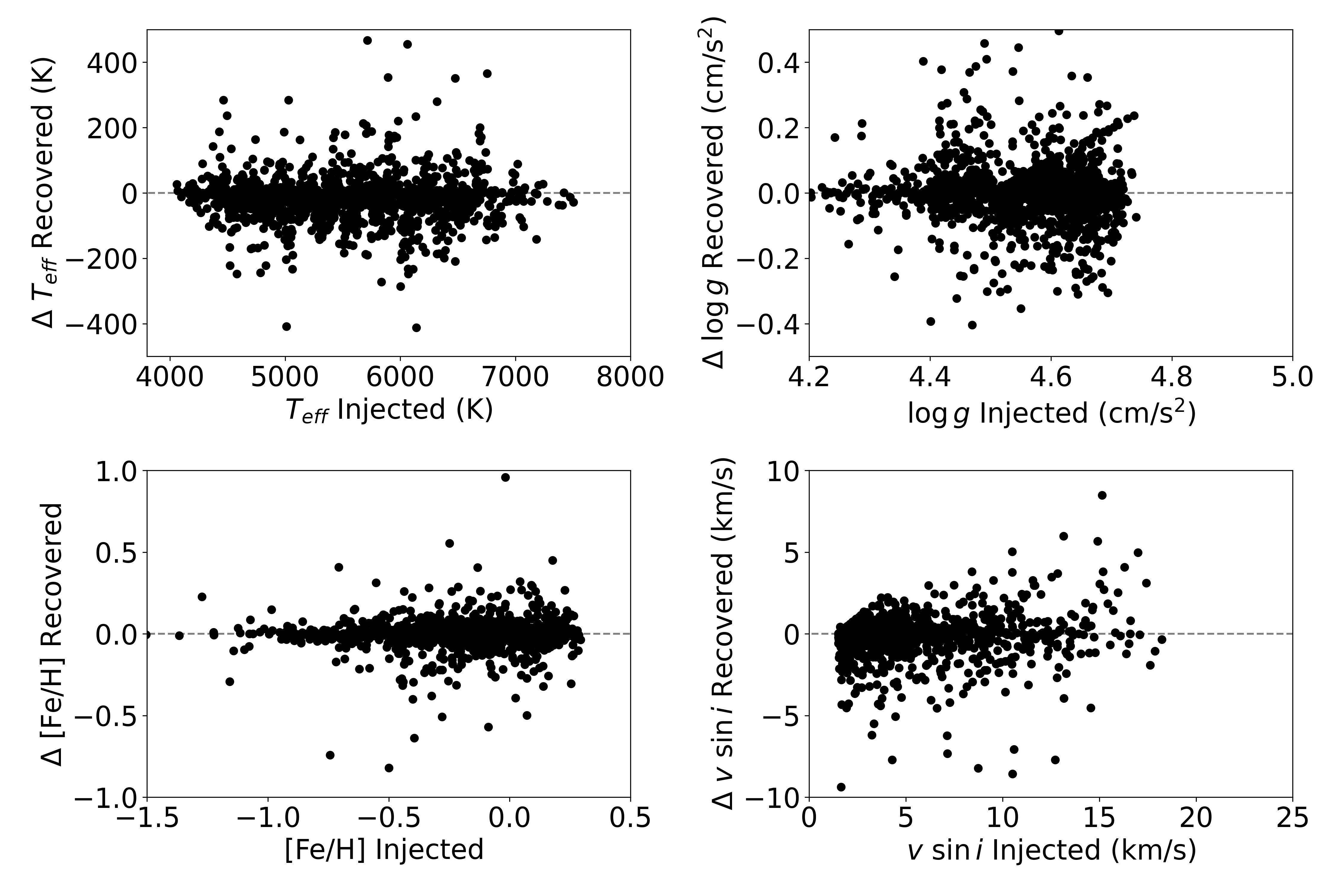}
    \caption{\tw{The difference between the recovered traditional stellar spectra parameters (\teff, \ \logg, \ [Fe/H] and \vsini) from the synthetic spotted spectra fitted with both a spotted and non-spotted model of the stellar spectra against the injected parameters of the synthetic spectra (spotted model non-spotted model recovered parameter). We identify scatter introduced to each of the stellar parameters when fitting spotted spectra with a non-spotted model of the stellar atmosphere.}}
    \label{fig:recov_dif}
\end{figure*}

We now identify systematic effects in the recovered parameters when we fit the spotted spectra with an incorrect model of non-spotted spectra. \tw{The difference between the recovered parameters fitted with a spotted and non-spotted model of the stellar atmosphere are shown in Figure \ref{fig:recov_dif}.} A consistent scatter is introduced on each parameter when a non-spotted model is used to perform inference on a spotted spectrum. We calculate each parameter's average bias and scatter to quantify the effect. The injected parameters are separated into 10 bins, and we take the median and median absolute deviation of the difference between the spotted and non-spotted model's inferred parameters for each bin. We take the median as a measure of the average bias and the median absolute deviation as a proxy for the scatter.

\begin{figure*}
    \includegraphics[width=\textwidth]{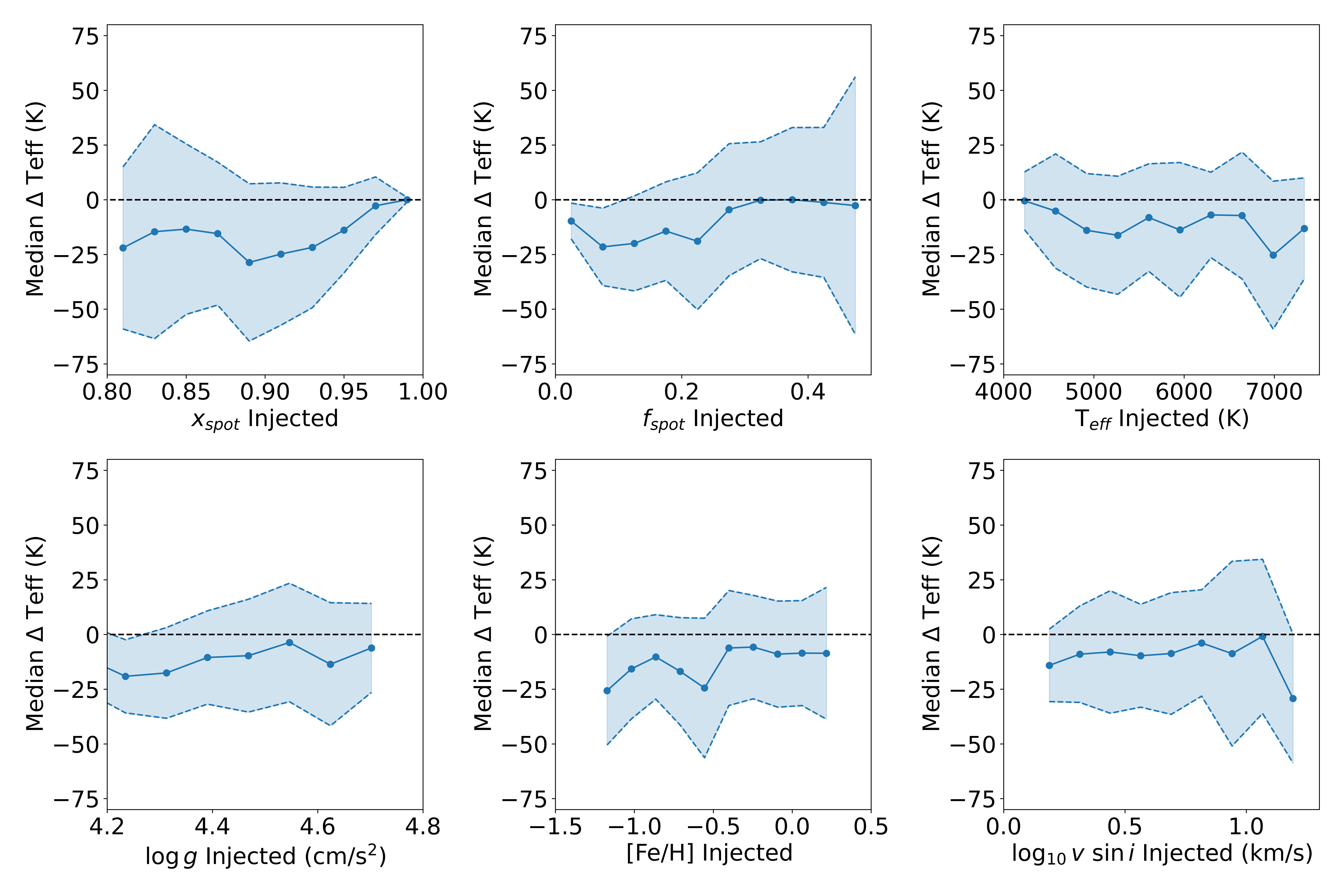}
    \caption{Bias introduced to \teff\ (blue) when fitting spotted spectra with a non-spotted model against injected parameters of synthetic spectra. Each injected parameter is binned into ten bins across the range of injected parameters. The median and median absolute deviation of the difference between the spotted and non-spotted recovered \teff\ ($\Delta$\teff) are then calculated in each bin.
    Scatter points show the median $\Delta$\teff\ for each bin in injected parameters. Filled areas show one maximum absolute deviation above and below the median value and dashed lines indicate the edge of this range.
     Inference of \teff\ with a non-spotted model injects random scatter on average of the scale of $\sim$50K and introduces a consistent bias of order $\sim$-25K for spectra with significant spot coverage.
    }
    \label{fig:res_teff}
\end{figure*}

In Figure \ref{fig:res_teff} we show the effect of the injected parameters on the stellar spot spectra through the difference between the recovered spot and non-spot model \teff. Fitting a spotted spectrum with a small \xspot with a non-spotted spectrum introduces a consistent bias to the inferred \teff\ of about $-25$\,K: a non-spotted model tends to underestimate the true effective temperature of a spotted spectrum. A scatter is also introduced \teff\ on the scale of $\sim$50K for spectra with significant spot coverage (low \xspot\ and large \fspot). The other injected parameters do not appear to have any strong correlations or effects on the recovered non-spot \teff. Their median values are zero, and MAD appears consistent at $\sim$25K.

\begin{figure*}
    \includegraphics[width=\textwidth]{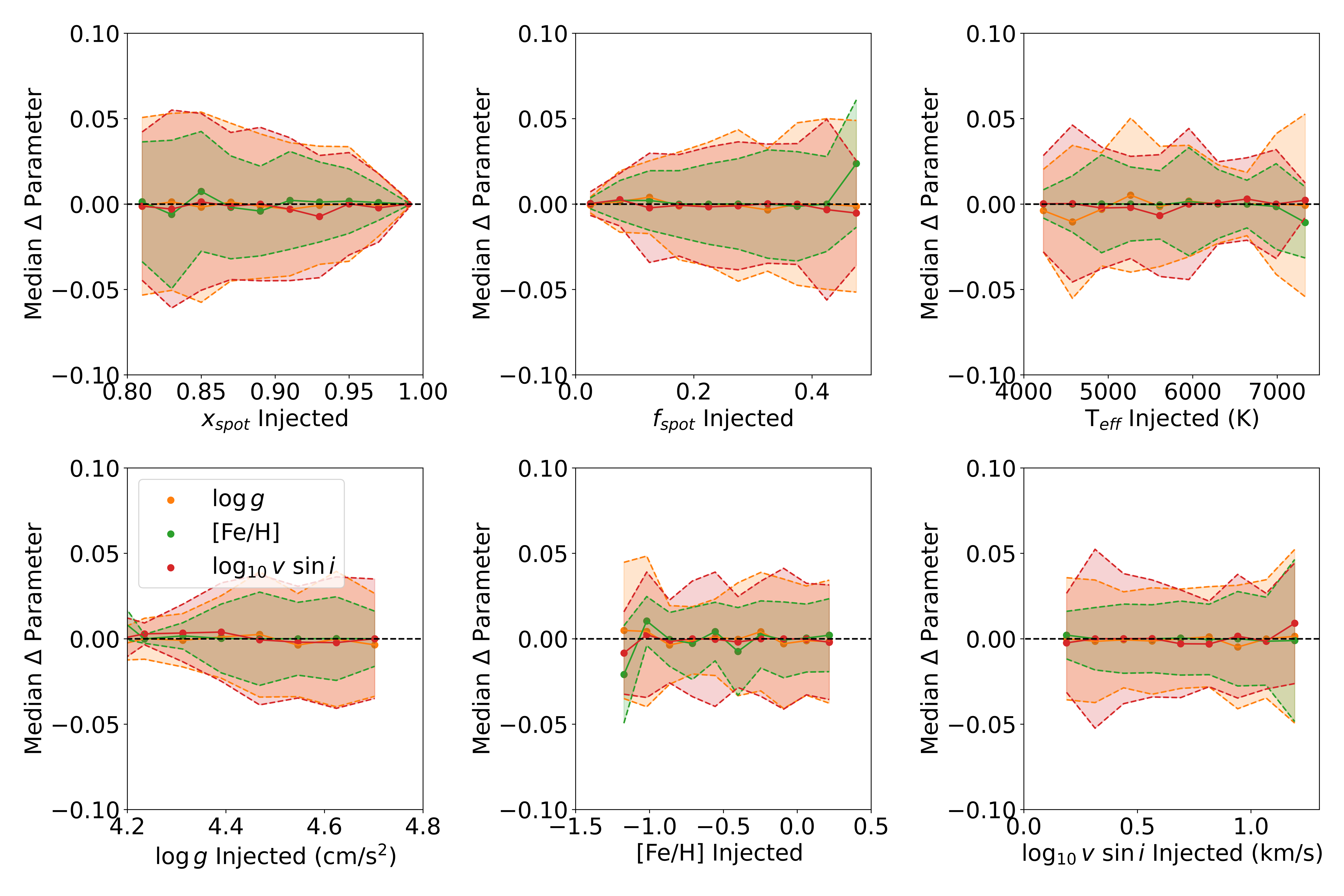}
    \caption{Bias introduced to \logg  \ (orange), [Fe/H] (green) and $\log$ \vsini\ (red) when fitting spotted spectra with a non-spotted model against injected parameters of synthetic spectra. \tw{Here we have plotted $\log$ \vsini\ as the grid we perform interpolation over is distributed uniformly in $\log$ \vsini\ and to make direct comparisons to the other stellar parameters ([Fe/H] and \logg). Each injected parameter is binned into ten bins across the range of injected parameters.} The median and median absolute deviation of the difference between the spotted and non-spotted recovered parameters ($\Delta$ parameter) are then calculated in each of these bins.
    Scatter points show each bin's median $\Delta$ parameter in injected parameters. Filled areas show one maximum absolute deviation above and below the median value and dashed lines indicate the edge of this range.
    Fitting spotted spectra with non-spotted models does not introduce a bias to the inferred parameters though it does introduce a scatter. This scatter increases with decreasing \xspot\ and increasing \fspot\ to a maximum median absolute deviation on the order of $\sim$0.05 for \logg, [Fe/H] and $\log$ \vsini \ - which corresponds to a maximum scatter on \vsini \ of $\sim$2 km/s.
    The scatter is otherwise approximately constant for all other parameters, and on average $\sim 0.025$ \ - which corresponds to an average scatter on \vsini \ of $\sim$ 1 km/s }
    \label{fig:res_full}
\end{figure*}

Figure \ref{fig:res_full} shows the effects of fitting spotted spectra with a non-spotted model on \logg\ (orange) and $\log$ \vsini\ (red), respectively.
There appears to be no statistically significant bias introduced to both of the inferred parameters as the median of each bin of injected parameters is consistently about zero. However, a consistent scatter is introduced to both parameters. The MAD of $\Delta$ \logg\ and $\Delta \log$ \vsini\ in each injected parameter bin have an average value of $\sim$0.025 dex - corresponding to an average scatter on \vsini\ of $\sim$ 1 km\,s$^{-1}$. The scatter peaks for both recovered parameters at $\sim$0.05 dex for stars with significant spot coverage - which corresponds to a maximum scatter on \vsini\ of $\sim$ 2 km/s. The scatter on recovered \vsini\ and \logg\ is otherwise constant with the other injected parameters.

The effect of fitting spotted spectra with a non-spotted model is significant in the recovery of metallicity. This is seen in Figure \ref{fig:res_full} (green), where we compare the recovered [Fe/H] with a spotted and non-spotted model of the stellar atmosphere against the injected parameters of our spotted spectra. This process does not introduce a bias to the inferred metallicity of the spectra but does introduce a significant scatter to the recovered value, representing an intrinsic `minimum floor' of systematic uncertainty if the effects of spots are not included (see Section~\ref{sec:discussion}).

\tw{The scatter introduced to [Fe/H] by fitting spotted spectra with a non-spotted model increases with injected \fspot. As \fspot\ approaches 1, the MAD of $\Delta$[Fe/H] reaches a maximum of about 0.05 dex.} Comparatively, as \xspot\ decreases, so does the MAD of $\Delta$[Fe/H], peaking again at 0.04 dex. As the scatter in the other injected parameters is relatively constant, there is no significant relation between the other spectral parameters and $\Delta$[Fe/H]. The introduced scatter in [Fe/H] is dominated by the spot parameters of spectra.

\begin{figure}
    \includegraphics[width=0.5\textwidth]{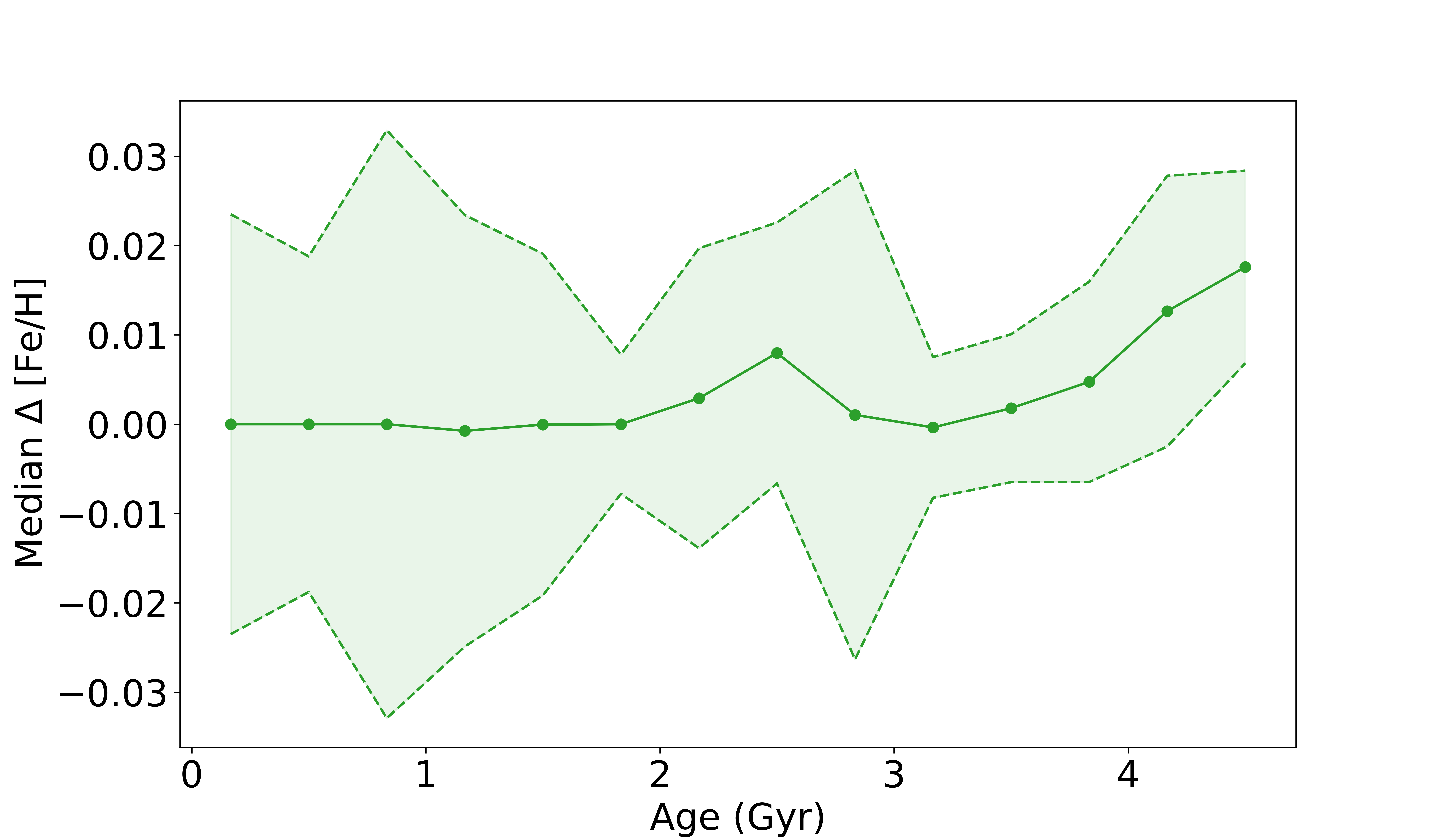}
    \caption{Bias introduced to [Fe/H] when fitting spotted spectra with a non-spotted model against the age of the model used to generate synthetic spectra. Age is binned into ten bins across the range of injected parameters. The median and median absolute deviation of the difference between the spotted and non-spotted recovered [Fe/H]\ ($\Delta$[Fe/H]) are then calculated in each of these bins.
    Scatter points show the median $\Delta$[Fe/H] for each bin in injected parameters. Filled areas show one maximum absolute deviation above and below the median value and dashed lines indicate the edge of this range.
    We find that the introduced scatter is greatest for stars younger than $\sim$ 2Gyr is $\sim$0.02 while the bias is 0 for stars in this age range.
    The scatter decreases for older stars ($>$3Gyr) to $\sim$0.01 but the median $\Delta$[Fe/H] increases with increasing age.
    This is like the result of the small number of stars in our sample in this age range rather than indicative of a trend.}
    
    \label{fig:age_mad}
\end{figure}

\section{Discussion}
\label{sec:discussion}

The results in Section~\ref{sec:results} indicate that using a non-spotted model to fit spotted stellar spectra introduces a systematic bias of up to $-25$\,K in effective temperature and no substantial bias in other parameters. In their study of fitting a spotted stellar model to APOGEE spectra of members in the Pleiades and M67, \citet{cao_starspots_2022} find a systematic 0.1 dex enhancement in observed [Fe/H]. The lack of bias we find here could be attributed to different stellar populations of stars (e.g., some stars are biased in one direction, but in our population, that effect is mitigated by biases in the opposite direction). We find that the effects of model mismatch (i.e., using a non-spotted model to fit a spotted spectrum) can also introduce a scatter (measured by median absolute deviation) of about 50 K in effective temperature and 0.05 dex in other parameters. If we assume that the spot model we adopt is representative of reality, then these scatter values would represent a minimum systematic uncertainty in these parameters if the wrong model (a non-spotted model) is used. These deviations are comparable to the typical random uncertainties reported by the APOGEE survey \citep[150K, 0.13 dex and 0.1 dex;][]{hegedus_comparative_2022}, although these random uncertainties will vary with signal-to-noise.

Systematic uncertainties (like model mismatches) will dominate in high signal-to-noise ratios, and the level of scatter we find in metallicity (0.05 dex) is comparable to the effects of radiative levitation, atomic diffusion \citep{onehag_abundances_2014}, and magnetic broadening of absorption lines \citep{spina_how_2020}. Unlike these effects, which can in part be mitigated through parameterisation with other stellar parameters, accounting for stellar spots requires a model that explicitly predicts their contribution to the emergent spectrum. This scatter in [Fe/H] is significant as it is of the same order as the precision of spectroscopic inference of metallicity. In particular, a differential analysis of two Solar twins might report abundance uncertainties on the level of 0.01-0.02 dex. While the two stars are selected to be extremely similar in order to mitigate systematic effects, those two stars could have very different coverages of stellar spots, which would introduce a systematic uncertainty floor. 

\subsection{Imperfect models}

The results we show here are limited in their applicability. When generating the mock data, only a fraction of randomly drawn stellar parameters could be used to synthesise spectra, either because of limitations of stellar isochrones, the spectral grid, or limits in the procedure in estimating an appropriate rotational velocity and Rossby number. We also limit the ages of the stellar sample to 4.6 Gyr - the maximum ages of both the models used to determine the convective turnover timescale and grid of rotational periods set by observations. As a result, our sample is limited to relatively young stars, and there are hints of a bias in injected parameters towards higher \fspot. We have extensively probed the region where the effect should be most prevalent in terms of the scatter it introduces, but this is not intended to be a complete and representative population of main-sequence stars. The quantitative results may not be perfectly accurate for some regions of the HR diagram\rev{, and might vary with photopshere geometry}. However, by assuming spots are present everywhere across the main sequence, our analysis shows where the consequential effects are most or least prevalent. 

The treatment of stellar spots in this work requires some discussion. Stellar spots are highly complex regions on the surface of stars. The position of spots relative to the observer, their temporal evolution, and the inherent magnetic activity and faculae surrounding stellar spots, would all introduce complexity to the emergent spectra from these regions. The spotted model employed in this work is a first-order approximation of the average effect of spots on stellar spectra. 
The functional form of the temporal evolution of the stellar spots in stars other than the Sun is not well known. For a given \fspot\, we could assume that \xspot\ varies on some periodic or temporal scale, even if we don't know the functional form of that variability. In this scenario with our model, \xspot\ is drawn from a uniform prior, which implicitly assumes that we are observing the star at some random time. This modelling of \xspot\ is relatively crude since, in principle, \xspot\ could vary as a function of other stellar parameters.

\tw{Investigations of the evolution of fractional spot coverage of stars is a developing field.
For example, recent works have shown an enhancement in \fspot\ for stars undergoing core-envelope recoupling \citep{cao_core-envelope_2023}.
For this reason, our results are only indicative rather than prescriptive. Applying this model to more stars APOGEE samples and time series spectroscopic observations of stars could elucidate the relationship between the parameters.}

\citet{cao_starspots_2022} suggest that young, magnetically active stars - stars with Rossby numbers < 0.4 - have \fspot\ greater than > 0.1, saturating at \fspot\ $\sim$ 0.3, with significant scatter, when $R_o$ < 0.2. 
There is also a significant scatter in \fspot\ for these stars. The use of the Rossby number to reflect the magnetic/spot activity of stars should be treated with some care. The Sun expresses periodic evolution of its magnetic activity (time scale on the order of decades) and stellar spot expression (time scale on the order of years). The range of fractional spot coverages we observe in the Sun is on the order of [0, 0.12] without variations in the Rossby number. As a result, we draw the injected \fspot from relations with Rossby number and add a random scatter drawn from a Gaussian distribution with a standard deviation of 0.1.

Employing a non-spotted spectra model to fit spotted spectra can introduce a scatter to recovered parameters, but fitting a spotted model to non-spotted spectra has little to no effect on the recovered parameters.
We recommend that a spotted model, if only as simple as the one used in this work, will consistently recover stellar parameters better than a non-spotted model while also providing a measure of the spot parameters of stars.

\subsection{When should a spotted model of the stellar atmosphere be employed?}

\tw{The scatter introduced to the recovered stellar parameters increases with fractional spot coverage.
Fractional spot coverage is inversely related to the rotation rate of stars through $R_o$.
Further, the rotation rate of stars decreases with time, owing to magnetic braking.
As a result, the fractional spot coverage of stars is expected to decrease with age.}

\tw{We can probe when the scatter introduced to the recovery of stellar parameters by stellar spots is most prominent by calculating the scatter in $\Delta$[Fe/H] with bins of age. In Figure \ref{fig:age_mad}, we show the bias and scatter introduced to $\Delta$[Fe/H] with respect to stellar age. The introduced scatter is greatest for stars younger than $\sim$ 2Gyr is $\sim$0.02, while the bias, measured through the median, is zero for stars in this age range.
The scatter decreases for older stars ($>$3Gyr) to $\sim$0.01, but the median $\Delta$[Fe/H] appears to increase with increasing age.
The increase in the median value is most likely not indicative of a trend and rather the result of the low number of stars in the larger age bins.}

\tw{The trends that we identify in this work are only qualitative - though they do allow us to make recommendations for future work.
Our results indicate that fitting the spotted spectra of a star with a non-spotted model when \fspot> 0.1 will introduce a scatter to bias the recovered parameters. We suggest using a spotted model if a star is significantly photometrically variable due to stellar spots.
\citet{mcquillan_rotation_2014} calculated the rotation periods of low-mass main-sequence stars that are photometrically variable due to stellar spots.
They were able to determine the rotation rates of stars across a wide mass range (0.6 < $M_\odot$ < 1.1) at multiple points along the main sequence.
These stars must therefore express stellar spots and may have the measured stellar parameters influenced by the effect we identify in this work.
They estimated that $\sim$23\% of main-sequence stars exhibit definite rotational modulation from stellar spots, a lower bound due to observational effects.
We, therefore, believe at least 1/4 of the main sequence stars may be affected by this bias.}

\section{Conclusions}
\label{sec:conclusions}

Here we have shown that stellar spots can introduce measurable systematic bias and variance to inferred stellar parameters when a non-spotted model is used. The results demonstrate that spectra with strong spot features can introduce a scatter in inferred metallicity of order 0.05 dex. This emphasises the need for caution when performing spectroscopic analysis on stars with visible spots, particularly young, fast-rotating stars. Our findings highlight the importance of incorporating the effect of spots into spectroscopic models to ensure accurate and precise results. 

The magnitude of this effect is comparable to others that plague stellar spectroscopy, including atomic diffusion, radiative levitation, and non-local thermodynamic equilibrium. However, the impact of this effect will vary depending on the scientific context. Turn-off ages of clusters are likely to be only minimally impacted, as the metallicity bias for old, slowly rotating stars is less than 0.01 dex. In contrast, a systematic error floor of 0.05 dex caused by spots on the main sequence would critically limit the capacity of strong chemical tagging \citep{casamiquela_impossibility_2021}. Similarly, star spots could limit any inferences from differential abundance analyses of Sun-like stars, where the typical reported uncertainty is 0.01-0.02 dex \citep[e.g.,][]{melendez_18_2014,nissen_high-precision_2015, reggiani_first_2016, maia_revisiting_2019, liu_detailed_2020, nissen_high-precision_2020, spina_chemical_2021}. While we have focused on the impact on overall metallicity and not on individual abundances, it will be important to examine these effects more closely at a per-element level. These results provide valuable insights for future studies on stars and their properties and underscore the need for continued research on the impact of spots on spectroscopic inference.

\section*{Acknowledgements}

\rev{We thank the reviewer for their constructive feedback which helped improve the paper.} We thank Prof. Ilya Mandel for his helpful feedback and our fruitful discussions.  \rev{A.~R.~C. thanks Jon Holtzman and Adam Wheeler for helpful discussions.}
A.~R.~C. is supported in part by the Australian Research Council through a Discovery Early Career Researcher Award (DE190100656). Parts of this research were supported by the Australian Research Council Centre of Excellence for All Sky Astrophysics in 3 Dimensions (ASTRO 3D) through project number CE170100013.

\section*{Data Availability}
The data and models underlying this article are available upon request to the corresponding author.

\bibliographystyle{mnras}
\bibliography{references}
\bsp	
\label{lastpage}
\end{document}